# Magnetoresistance and collective Coulomb blockade in super-lattices of ferromagnetic CoFe nanoparticles

R. P. Tan[1,+], J. Carrey[1,*], C. Desvaux[2], L.-M. Lacroix[1],
P. Renaud[2], B. Chaudret[3], M. Respaud[1,*]

[1] Université de Toulouse; INSA, UPS; LPCNO, 135, av. de Rangueil, F-31077 Toulouse, France
CNRS; LPCNO, F-31077 Toulouse, France
[2] Semiconductor Products Sector, Freescale, le Mirail B.P. 1029, 31023 Toulouse Cedex, France
[3] Laboratoire de Chimie de Coordination-CNRS, 205 rte de Narbonne, 31077 Toulouse cedex 4, France

**Abstract :**

We report on the magneto-transport properties of chemically-synthesized magnetic artificial solids consisting of millimetre-size super-lattices of CoFe nanoparticles (NPs) separated by a thin organic insulating layer. Electrical measurements highlight the richness of the interaction between transport and magnetic field in three-dimensional networks of magnetic NPs, especially in the Coulomb blockade regime: (i) Resistance-temperature characteristics follow $R = R_0 \exp[(T_0/T)^{1/2}]$, as generally observed in NP arrays displaying charge or structural disorder. (ii) Low-temperature current-voltage characteristics scale according to $I \propto [(V-V_T)/V_T]^\zeta$ with $\zeta$ ranging from 3.5 to 5.2. For a sample with a very large size distribution of NPs, a reduced exponent down to $\zeta = 1$ is found, the origin of which remains unclear. (iii) A large high-field magnetoresistance displaying a strong voltage-dependence and a scaling versus the magnetic field/temperature ratio is observed in a limited temperature range (1.8 K-10 K). The most likely interpretation is related to the presence of paramagnetic centres at the surface or between the NPs. (iv) Below 1.8 K, concomitantly to the collapse of this high-field MR, a low field inverse tunnelling magnetoresistance grows up with a moderate amplitude not exceeding 1%. (v) Below a critical temperature of 1.8 K, abrupt and hysteretic transitions between two well-defined conduction modes – a Coulomb blockade regime and a conductive regime – can be triggered by the temperature, electric and magnetic fields. Huge resistance transitions and magnetoresistance with amplitude as high as a factor 30 have been observed in this regime. We propose that these transport features may be related to collective effects in the Coulomb blockade regime resulting from the strong capacitive coupling between NPs. They may correspond to the soliton avalanches predicted by Sverdlov *et al.* [Phys. Rev. B 64, 041302 (R), 2001] or could also be interpreted as a true phase transition between a Coulomb glass phase to a liquid phase of electrons. The origin of the coupling between magnetic field and transport in this regime is still an open question.

Spin-polarized transport processes, 72.25.–b

Electrical conductivity of disordered solids, 72.80.Ng

# I. Introduction

In the nanotechnology-based world which is predicted to emerge, magnetic nanoparticles (NPs) could play a major role. In the field of bio-nanotechnologies, they are already used as contrast agents in magnetic resonance imaging and for magnetic separation, but are also promising materials for magnetic hyperthermia applications or drug targeting [1]. With respect to magnetic recording, the increase of storage density requires the optimization of high-anisotropy nanomaterials [2]. Finally, using magnetic NPs with well-controlled properties may open new opportunities in spintronics. For instance, in the field of spin-transfer, this could lead to a strong reduction of the injection current for magnetization switching and/or radio-frequency precession [3,4]. Moreover, the interaction between Coulomb blockade and magnetic field also leads to a very rich physics opening the way to innovative devices [5, 6, 7, 8].

Chemical synthesis is a preferred way to elaborate large quantities of well-controlled nano-objects that can be further used as elementary building blocks. In the case of magnetic NPs, the optimisation of the synthesis procedure allows now a fine control not only of their size, but also of their surface chemistry and anisotropy. For instance, polymer coated Co or Fe NPs can display magnetization enhancement similarly to free NPs studied under high vacuum [9,10]. The possibility offered by the chemistry to tune NP properties could be of great interest for fundamental studies and applications in spintronics. However, the study of the effects mentioned above require the connection of single chemically-synthesized NPs, a challenging technological task which has not yet been successful.

On the other hand, recent progresses in NP synthesis have permitted the elaboration of artificial solids consisting of well-defined networks of NPs. Several techniques are available, such as layer-by-layer deposition [11], crystallisation [12], or direct self-assembling during synthesis [13]. Such systems are models for the investigation of collective physical properties (electric, magnetic, optical, vibrational, …). In the framework of spintronics, experimental investigations of large arrays of chemically-synthesized magnetic NPs are rather scarce. After pioneer works by Black *et al.* on Co NPs [14], several studies have focused on magnetic oxide NPs for which magneto-transport properties are mainly governed by unsaturated spins at the NP surface [15].

Theoretically, arrays composed of magnetic NPs separated by an insulating tunnel barrier may display both tunnel magnetoresistance (TMR) and Coulomb blockade properties. TMR is due to the fact that the electron tunnel rate between two spin-polarized particles depends on the angle between their magnetization. In the simplest model, assuming that electrons tunnel sequentially between NPs and that the NPs have random anisotropy axes, TMR ratio is given by [16]:

$$\text{TMR} = (R_{AP} - R_P)/R_P = 2P^2/(1+P^2), \qquad (1)$$

where $R_{AP}$ is the resistance at the coercive field of the array, $R_P$ the resistance when the NPs are magnetically saturated, and $P$ their spin polarisation. Even with a complete spin polarisation of the NPs ($P = 1$), the maximum TMR value which can be reached with this mechanism is 100 %. This is due to the fact that an anti-parallel alignment of the NP magnetization is never reached in NP arrays, oppositely to magnetic tunnel junctions, in which TMR can in principle be infinite using perfectly spin-polarised materials. When the NPs are composed of the standard 3d metals (Fe, Co, Ni) and their alloys, for which the maximum spin polarisation is about 50 % [17], the maximum TMR ratio expected is thus 40 %. Experimental TMR values never exceed a few tens of percent [14,18,19,20]. Higher values are observed when transport between NPs occurs via other mechanisms like co-tunnelling (see below).

Coulomb blockade in arrays of NPs is due to the electrostatic charging energy $E_C$ necessary to add a new charge to a NP with a finite capacitance or to move a charge between neighbouring NPs [21,22,23]. $E_C$ depends on the self-capacitance of the NPs and on their mutual capacitance. A well-known consequence of the Coulomb blockade in arrays of NPs is the apparition of a gap in the current-voltage [$I(V)$] characteristics of the arrays below a temperature $T$ such as $k_B T << E_C$. Above this gap, numerical studies predict that $I(V)$ characteristics follow at $T = 0$:

$$I \propto [(V-V_T)/V_T]^\zeta, \qquad (2)$$

where $V_T$ is the threshold voltage above which conduction occurs [21,22,24]. This power-law is related to the progressive opening of conduction channels, so $\zeta$ is related to the dimensionality of the array. $\zeta = 1$ and $\zeta = 5/3$ have been calculated for 1D and 2D arrays of disordered NPs respectively. No theoretical value has been calculated for 3D arrays, even if larger values than for the 2D case are expected. For voltages well above $V_T$, the differential conductivity should follow a linear regime [23]. Besides, the evolution of the low-bias resistance as a function of temperature in an array of NPs is expected to follow [25]:

$$R = R_0 \exp[(T_0/T)^\nu], \qquad (3)$$

where $R_0$ is the high-temperature resistance, $T_0$ the activation temperature for charge transport and $\nu$ an exponent generally found equal to 0.5 [26, 27, 28].

Arrays of NPs in the Coulomb blockade regime can also present other peculiar properties due to the capacitive coupling between NPs. For instance, a frustrated and non-ergodic phase called "Coulomb glass" is expected to appear in such systems, resulting from the interaction between electrostatic forces and charge disorder in the array [29]. In this state, slow relaxation, aging and hysteretic phenomena are expected. Collective effects such as avalanches have also been predicted in the strong coupling regime [30]. However, these effects have so far not been observed in arrays of NPs but only in granular metals [31,32] and doped semiconductors [33], which are also arrays of localised interacting electrons.

The interplay between Coulomb blockade and spin dependent tunnelling or a magnetic field leads to various effects, which have been recently reviewed by Yakushiji *et al.* [34] and Seneor *et al.* [5]. First, in arrays of NPs, when co-tunneling occurs, *i.e.* when an electron tunnels to a distant NP via other ones, the TMR value is enhanced. This enhancement, which depends on the order of co-tunneling, has been for instance experimentally observed in insulating Co-Al-O granular films [7,35]. Second, the so-called "magneto-Coulomb effects", due to a shift of the magnetic NP electrochemical potential by a magnetic field, can lead to very large MR ratios in the Coulomb blockade regime, especially when the applied voltage is close to $V_T$ [36,37]. This last effect has so far only been theoretically studied and experimentally reported in devices where a single NP is measured.

We recently reported the observation of several novel transport features in three-dimensional super-lattices of CoFe NPs [38,39,40,41]: i) a large high-field MR which could reach up to 3000 % ii) the presence of hysteretic step jumps between two states in the *I*(*V*) characteristics of samples in the Coulomb blockade regime iii) in the latter samples, the possibility to induce the transition between the two states by a magnetic field iv) a low field MR with a complex shape. Thus, these super-lattices display several mechanisms of MR which can obviously not be explained by the standard mechanisms of MR presented above. Moreover, we observed an original coupling between the Coulomb blockade and the magnetic field. In this article, we report on the properties of six samples which were extensively measured and analysed. We will stress on the common points and the differences, so that a global view of their properties can be drawn. The paper is organised as follows. In the second section, we describe the main characteristics (synthesis and magnetic properties) of the samples. The third section is devoted to the description of transport measurements within the different regimes. Finally these results will be discussed in section IV.

II. Synthesis and characterisations.

An organometallic approach consisting on the co-decomposition of two precursors in the presence of stabilizing agents was used to synthesize the FeCo NPs super-lattices. The variation of the ligands gave different materials referred as 1 OA, 1 OA+1 SA, 2 OA in Table I (see below). The principle of the synthesis has been reported in details elsewhere [13,42]. Briefly, one equivalent of the cobalt precursor [Co($\eta^3$-$C_8H_{13}$)($\eta^4$-$C_8H_{12}$)] was mixed in toluene under inert atmosphere with 2 equivalents of the iron precursor [Fe(CO)$_5$] in the presence of ligands : i) 1 equivalent of oleic acid (1OA, Sample D); ii) 2 equivalents of oleic acid (2 OA, samples A and E); iii) 1 equivalent of oleic acid and 1 equivalent of stearic acid (1 OA+1 SA, samples B and C). The reaction mixture was then heated to 150°C under 3 bars of dihydrogene for 2 days. At the end of the synthesis, the colloidal solution is dried under vacuum. The super-lattices were collected as small needles. The final composition of the obtained material was set to 60 % iron and 40 % cobalt.

When 1 eq. acid oleic (sample D) or a mixture of 1 eq. oleic acid and 1 eq. stearic acid (samples B and C) is used as stabilizing agents, very monodispersed 15 nm particles are obtained.

They spontaneously self-organize with a long range order (as visible on a microtomy cut of a sample [13]) to form super-crystals, with a fcc packing of the particles [see Fig 1(b)]. When 2 eq. of oleic acid are used (sample A and E), the particles are bigger (25 nm diameter) and less monodisperse. This leads to a lack of straightforward arrangement of the particles within the super-lattices, which thus consist of densely packed NPs with no long range spatial ordering [see Fig 1(a)].

Wide angle X-ray scattering and high resolution transmission electron microscopy (TEM) evidence no classical crystalline phase in the NP. Parallel electron energy loss spectroscopy (PEELS) analysis shows that NPs display an onion-like structure consisting of a cobalt-rich core surrounded by an iron-rich shell, itself surrounded by another cobalt-rich shell [21].

Magnetization characteristics on a powder composed of many super-lattices is shown in Fig. 2(a). The saturation magnetization $M_S = 163$ A.m$^2$.kg$_{FeCo}^{-1}$ at 2 K is lower than the bulk alloy ($M_S = 240$ A.m$^2$.kg$_{FeCo}^{-1}$). The segregated structure and amorphous character of the particle may be responsible for the low magnetization. A coercive field of 10 mT and a saturation field of 1.2 T are measured at 2 K. No oxidation is detected, either in wide angle X-ray scattering, energy electron loss spectroscopy, or in the magnetic measurements. Moreover, the bulk magnetization of FeCo is recovered after annealing, as the particles crystallize in the bcc alloyed phase. Fig. 2(b) displays the magnetization curves measured on a single super-lattice, with the magnetic field applied successively perpendicular and parallel to its longest axe. The coercive field is independent of the magnetic field orientation and equals 5 mT. The saturation field is slightly higher in the perpendicular direction (0.9 T) than in the parallel one (0.75 T).

Finally, sample X results from a synthesis involving 1 eq. oleic acid and 1 eq. stearic acid. Still, a problem in the manipulation or with the products purity led to a totally different material. The composition of the material was found to be 30 % iron and 70 % cobalt. The NPs were extremely polydisperse, with typical sizes ranging from 15 to 50 nm. Fig.1(d) and 1(e) show respectively a SEM micrograph of the sample X and a TEM micrograph of a super-lattice from the same synthesis after redispersion in THF. In spite of this polydispersity, the NPs were still compactly packed in millimetre-size super-lattices, which made this sample a good candidate to test the influence of structural disorder and of size distribution on the transport properties. As will be more evidenced below, the transport properties of this sample were indeed of particular interest.

As grown super-lattices were connected using Au wires and silver paint in a glove box under Ar atmosphere. The typical distance between the two contacts is 0.5 mm. The time to transfer the connected samples from the glove box to the inside of the cryostat was kept to a minimum (few tens of seconds) to prevent oxidation. Magnetotransport measurements were performed in a Cryogenic cryostat equipped with a superconducting coil using a Keithley 6430 sub-femtoamp sourcemeter. In these experiments, the magnetic field was applied perpendicularly to the current and to the longest axis of the super-lattice, with a constant sweep rate of 0.009 T/s.

III. Transport properties

A. Room temperature resistances and choice of the samples.

All the samples measured displayed at room temperature linear $I(V)$ characteristics. The room temperature resistances were measured in a range between 8 k$\Omega$ and 5 G$\Omega$. To explain this broad distribution, one possible hypothesis could be the contact resistance between the silver grains of the silver painting and the super-lattices. To test this hypothesis, we have also measured the room temperature resistances of super-lattices on which we evaporated a thin gold layer with a 20 µm gap. No effect was observed on the values of the resistances. The origin of the broad dispersion is probably inherent to the structural properties of the super-lattices themselves: a variation in the thickness of the organic tunnel barriers surrounding the NPs or microcracks inside the super-lattice could be at its origin. The room temperature resistances of the samples are summarized in Table I. Statistical conclusions on the resistance of the super-lattices cannot be deduced from this table. Indeed, as will be detailed below, the low-resistance super-lattices were the only ones to display hysteretic jumps in their $I(V)$ at low temperature. We thus connected inside the glove box many samples, checked their resistances *in situ* and measured extensively only the low-resistance ones. Some samples were also only briefly measured at low temperature, did not show any jump in their $I(V)$ and were as a consequence not further investigated. They are not included in Table I.

### B. Coulomb blockade properties

All the samples displayed a strong increase of their low-bias resistance when decreasing temperature combined with the apparition of non-linearity in their $I(V)$ characteristics. These are two features typical of Coulomb blockade in arrays of NPs. Fig. 3 shows the $R(T)$ characteristics measured on the samples. Given the large dispersion in their resistances, the voltage bias used to measure them varies from sample to sample and was chosen so the resistance could be still measurable at low temperature. These voltages are indicated in parentheses in the "$T_0$" column of Table I. In all the samples, $R(T)$s follow Equ. (3), with a coefficient $\nu$ equal to 0.5 for temperatures between 100 and 10 K (see Fig. 3). Samples D and X display below 10 K a noticeable deviation to this law with an additional enhancement of the resistance. The activation energies $T_0$ deduced from the linear slope between 10 and 100 K range from 13 to 256 K (see Table I).

Low-temperature $I(V)$ characteristics of arrays of NPs in the Coulomb blockade regime are expected to follow Equ. (2). As will be detailed in section D, some samples display below 1.8 K, avalanches in their $I(V)$ which cannot be analysed in this framework. Thus, we only analyse in this section the $I(V)$ characteristics measured at the lowest temperature possible just above the onset of avalanches. These $I(V)$s are summarized in Fig. 4. To deduce $\zeta$ and $V_T$ values from these data, we first plot the function $I.dV/dI$ versus $V$, which is linear above $V_T$ and allows one to determine $\zeta$. These plots are displayed in the insets of Fig. 4. Then, we estimate the $V_T$ value in a classical least-square fit of the $I(V)$ using $\zeta$ as a fixed parameter. Samples A and C do not display clear visible gap in their $I(V)$ characteristics so the $V_T$ values extracted from these fits are almost null. They also display a rather low $\zeta$ value compared to the other samples. This is due to the fact that Equ. (2) is in principle strictly valid at $T=0$, or in a range of temperature where $I(V)$ curvature remains constant, *i.e.* $V_T$ is just shifted when varying the temperature [24]. For samples A and C, this condition is not fulfilled, presumably because the Coulomb gap may open at an even lower temperature. Sample B, D and E display a clear gap in their $I(V)$ characteristics and physically reasonable values of $\zeta$ and $V_T$ are deduced (see Table I). However, it must be pointed out that there is a strong interdependency between $\zeta$ and $V_T$ in a least square fit. To illustrate this, we also display in Table I fit results where we arbitrary put $V_T$ to another value for which the fit is still very good (the chi-square value is increased by a factor of two and the two fits cannot be distinguished with the eye). These results show that the error bars on $\zeta$ is around 1.

We now analyse the evolution of the $I(V)$ characteristics as a function of the temperature. In the framework developed by Parthasarathy *et al.* [24], when the Coulomb blockade regime is fully established, it is expected that the $I(V)$s measured at various temperatures can collapse on a single curve when shifted by a temperature-dependant voltage. This analysis is presented for samples B and D in Fig. 5. For these samples, the shifted $I(V)$s collapse on a single curve in the range 1.7-3.7 K and 1.8-10.5 K, respectively (see Fig. 5(b) and 5(d)]. At higher temperatures, only the high voltage part of the $I(V)$ could collapse, but the non-linear part of the $I(V)$s at lower voltage do not superpose since the Coulomb gap is not fully established. Fig. 5(e) shows that the voltage used to shift the $I(V)$s varies linearly with the temperature. Similar behaviour was observed on 2D arrays of Au NPs [24].

Sample X has very peculiar $I(V)$ characteristics, which has to our knowledge never been reported in large arrays of NPs. At low temperature, above $V_T$, the $I(V)$ is linear, with an extremely narrow transition between the Coulomb gap and this linear part, meaning that all the conduction channels open very rapidly when the voltage is increased above $V_T$ [see Fig.4(f)]. In these conditions, it is not possible to deduce $\zeta$ from the $I(V)$, or it can alternatively be said that $\zeta = 1$. Fig. 5(f) displays the temperature-dependance of $I(V)$ characteristics, which illustrates the unusual behaviour at high voltage of this sample: all these $I(V)$s reach the same curve at high voltage, as if the sample was in a purely metallic state at high voltage, even at low temperature.

C. Collective Coulomb blockade and magnetic-field induced switching

Below a critical temperature of 1.8 K, the $I(V)$ curves of Samples A, C and X display original features with abrupt jumps from a highly resistance state to a more conductive one, and hysteresis when sweeping *V* up and down. These phenomena are only observed in the more conductive samples with a room temperature resistance not exceeding 400 kΩ. *I(V)* characteristics for sample A, C and X are displayed in Fig. 6(a), 6(d) and 7(a), respectively. In all cases, the jumps from the resistive state to the conductive one are correlated with a clear increase of the differential conductivity. The amplitude of these jumps measured at the lowest available temperature (1.5 K) are different from one sample to another, the maximum of which reaches 36 %, 120 % and 3000 % for sample A, C and X respectively. Applying a magnetic field modifies both the transition voltage and the shape of the hysteresis, as shown in Fig. 6(b), 6(e) and 7(b). However, this influence is not strictly similar in the three samples. In sample A, the hysteresis is not well established, since the current for $\mu_0 H = 0$ T oscillates telegraphically between the conductive and the resistive state, but with a higher probability to be in the resistive (conductive) state during the voltage sweep up (down) [see Fig 6(a) and 6(b)]. Upon application of a magnetic field, the hysteresis is shifted toward the left and its area progressively shrinks [see Fig. 6(b)]. In sample C, we were able to measure the evolution of the hysteresis in a small temperature range from 1.8 K to 1.5 K. Upon lowering the temperature, the poorly defined hysteresis at 1.76 K progressively grows up and becomes well-defined, with an abrupt jump on the sweep up and on the sweep down at 1.5 K [see Fig. 6(d)]. In this case, the application of the magnetic field slightly shifts the whole hysteresis toward the left [see Fig. 6(e)]. In sample X, a huge current drop appears in the sweep up, but no transition in the sweep down [see Fig. 7(a)]. In this case, the effect of the magnetic field is to shift the jump on the sweep up toward the left. It can be observed that the voltage shift is much larger in this sample than in sample C.

We also demonstrate that the transition between the two regimes can be triggered by the magnetic field in given conditions of applied voltages [see Fig. 6(c) and 7(c)]. Typically, a magnetic-field induced transition from the resistive state to the conductive one is obtained when the sample is polarized under a voltage slightly below the critical one. These transitions can be either reversible or irreversible, this (ir)reversibility being directly correlated to the width of the hysteresis of the $I(V)$. For instance, an irreversible transition on sample X is shown in Fig. 7(c). The sample is first placed in the resistive state at $V = 36$ V and the magnetic field is increased. For $\mu_0 H = 2.65$ T, the sample switches to the conductive state and remains in this state even when sweeping down the magnetic field. The critical field well matches the one that can be estimated from the $I(V)$ characteristics, *i.e.* between $\mu_0 H = 2.65$ T and $\mu_0 H = 4.6$ T for $V=36$ V [see Fig. 7(b)]. A reversible transition measured in sample A is shown in Fig. 6(c): under a polarisation of $V = 50$ V, the sample switches between the conducting and the resistive state for magnetic fields values around 1.5 T. The reversibility is in agreement with Fig. 6(b). Indeed, it shows that for $\mu_0 H = 0$ T, the hysteretic part starts above 50 V, and for $\mu_0 H = 3$ T, the hysteretic part ends around 50 V. This means that at $V = 50$ V, only one state is possible for $\mu_0 H = 0$ T (the resistive one), and one state is possible for $\mu_0 H = 3$ T (the conductive one), which thus allows one to switch reversibly from one state to the other by sweeping the magnetic field. In sample C, since the shape of the hysteresis evolves from a small width hysteresis at 1.8 K to a large width one at 1.55 K, both types of transitions are observed : the reversible one at 1.8 K and the irreversible one at 1.55 K [38]. This notion of (ir)reversibility depends both on the magnetic field and on the applied voltage. For instance, given the evolution of the $I(V)$ characteristics of sample C with the magnetic field at 1.55 K [see Fig. 6(e)] and assuming that the hysteresis keeps on shifting to lowest voltage linearly with the magnetic field, reversible transitions may be observed at $T = 1.55$ K at the condition to measure the sample at $V = 16.5$ V and with a maximum magnetic field around 35 T.

Finally, these abrupt transitions can also be observed upon sweeping up and down the temperature in between 1.5 K and 2 K. Fig. 7(d) displays the $R(T)$ hysteretic curves measured on sample X under a constant applied voltage, and a slow sweeping up and down of the temperature. When this voltage is 30 V, the sweeps up and down are superposed, confirming that the experiment is slow enough so that there is no lag between the temperature measured on our sensor and the temperature of the sample. When the voltage is 35 or 40 V, hysteresis and abrupt jumps appears in the $R(T)$ curves, with an hysteresis width around 300 mK. This behaviour is in good agreement with the $I(V)$ characteristics [see Fig. 7(e)]. When plotted in semi-logarithmic scale, it can be seen that the transition below 1.8 K goes with an abrupt increase of the transition voltage inducing a "no man's land" area in the upper right quarter of this graph. This graph indicates that this area can be crossed as a function of the temperature at the condition that the applied voltage is above 30 V, as demonstrated by the $R(T)$ measurements.

D. Magnetoresistance for $T > 1.8$ K

All the samples display a large high-field MR in the range 1.8-15 K. The maximum amplitude measured varies from sample to sample and is summarized in the column "high-field MR" of Table I. The highest MR value was observed in sample B and reached 3000 % [39]. All the samples present a similar evolution of the MR amplitude as a function of the temperature, which is summarized in Fig. 8(a) for samples B, C and D: the MR ratio increases when the temperature is lowered until an abrupt drop of its amplitude occurs below 1.8 K. We present now detailed results on the MR properties above 1.8 K of sample D, which are representative of all the samples. Detailed results on samples B and C in this regime can be found elsewhere [38,39]. In Fig. 8(b), a typical $R(H)$ measured on sample D is shown. $R(H)$ characteristics always have a butterfly shape and do not saturate at high magnetic field, even when a 9 T magnetic field is applied [39]. The butterfly shape is due to the fact that there is a lag of around 90 s between the variation of the magnetic field and the associated variation of resistance in the sample. To illustrate this point, we show in Fig. 8(c) an experiment during which the resistance of the sample is measured as a function of time when the magnetic field is swept from 0 to 2.6 T and then back. The times at which the magnetic field starts and stops its variation are indicated as vertical lines. During the sweep up, the resistance starts varying around 90 s after the sweep start. Similarly, the resistance stabilizes to its final value around 90 s after the magnetic field has stabilized to its final value (see Fig. 8(c) between $t = 200$ s and $t = 290$ s for instance). Thus, we emphasize that this hysteresis MR is an intrinsic dynamic effect of the super-lattices. In Fig. 8(d), MR curves measured at various temperatures ranging from 2.08 K to 13.5 K are plotted as a function of the ratio $H/T$. They superpose on a single characteristic, which indicates that the MR amplitude only depends on the $H/T$ ratio, although some deviations are observed at $T = 2.08$ K and $T = 2.66$ K. These deviations could arise from the fact that the curves are measured, in these cases, for temperatures very close to the transition one (collapse of the high-field MR (see Fig.8 (a)).

In Fig. 8(e), the voltage dependence of this high-field MR is shown. Three different methods of measuring this voltage dependence are compared. Blue dots correspond to the MR ratios deduced from individual $R(H)$ characteristics measured at various voltages. Black line and red open dots correspond to the MR values deduced from two $I(V)$ characteristics measured at zero field and 2.6 T. Black line originates from $I(V)$ characteristics measured using a constant current range, which explains the noise at low voltage due to an imprecision in the measurements of low currents. For red dots, the current at various voltages has been manually measured using the best measurement range for each point. As expected, the two last methods give identical results at high voltage. However the MR values deduced from $R(H)$ measurements are always 20 to 25 % lower than the values deduced from *I*(*V*) characteristics. This is well explained by the dynamical experiments shown in Fig. 8(c): during *R*(*H*) measurements, the maximum MR value is not reached, since the magnetic field is swept back before the resistance has reached its final stabilized value. From Fig. 8(c), it is deduced that the difference between the dynamical MR and the static MR should be in the range 20-23 %, in good agreement with what is indeed observed in Fig. 8(e).

E. Magnetoresistance for *T* < 1.8 K

Below 1.8 K, the high-field MR collapses in all the samples. Some of them display in a restricted range of applied voltage the magnetic-field induced switching between two conducting states, as detailed in section D. In addition, all the samples also display a small amplitude MR observed in a broad range of applied voltage. In Fig. 9 the MR properties of sample B are shown. In Fig 9(a), the *R*(*H*) characteristic measured up to 8.8 T evidences that a small-amplitude high-field MR is still present. However, its characteristic is radically different from the one observed above 1.8 K: it does not display any hysteresis, its curvature is rather downward than upward and its amplitude is much smaller. The voltage dependence of this high-field MR has not been studied. In Fig. 9(b), 9(c) and 9(d), *R*(*H*) characteristics measured at different voltages and at a lower field of 2.7 T are shown. They have a typical shape of inverse tunnel MR with two peaks at $\mu_0 H = \pm 0.1$ T and a saturation field around 1 T. The tunnel MR ratio is thus defined using the resistance at the peak and the resistance measured at 1 T. It evolves from -1.8 % at 90 V to -0.3 % at 200 V in sample B. Superimposed on this tunnel MR, a positive MR peak centred around zero is observed on all the samples. Its amplitude ranges from 0.05 % to 0.1 % in the various samples, and is independent of the applied voltage. When the sample is let relaxing at zero magnetic field during one or two minutes before a MR measurement, this central peak appears more evidently, displaying an enhanced amplitude during the first sweep of the magnetic field. Indeed, during this relaxation, the resistance of the sample slowly increases, reaching a level of resistance which is then never reached again during the MR measurement. This enhanced peak at the first sweep is noticeable in Fig. 9(b), 9(c) and 9(d) for sample B. It is not always trivial to distinguish this peak from the inverse tunnel MR characteristic presented before. In sample C, this distinction is easier and is now presented (see Fig. 10). In Fig. 10(a), the *I*(*V*) characteristic of this sample at 1.5 K is recalled with vertical lines indicating the various voltages at which subsequent MR measurements have been performed. For an applied voltage of 0.1 V, the inverse tunnel MR and the central peak are both visible, with a MR ratio of -0.8 % for the inverse tunnel MR (see Fig. 10 (b)). For *V* = 16 V, this MR reduces to -0.16 % and for *V* = 20 V no inverse tunnel MR is observed and only the central peak remains [see Figs. 10 (c), 10 (d) and 10 (g)]. Interestingly, at 1.5 K, two conduction states are possible at *V* = 18 V allowing the investigation of the low field MR in the two conduction states. These measurements are shown in Fig. 10(e) and 10(f), and are labelled as "low state" and "high state" for the resistive and the conductive state, respectively. The inverse tunnel MR is only observed in the resistivity state and completely disappears in the conductive state.

IV. Discussion

We first discuss the classical Coulomb blockade properties presented in section III.B. It should be first recall that there is so far no experimental study of Coulomb blockade properties in 3D super-lattices of NPs. However $\zeta$ values greater than the ones measured in 2D arrays of NPs are expected. $\zeta$ values ranging from 2.01 to 2.27 were reported in 2D arrays consisting of a single layer of NPs deposited on a substrate [24,26]. Thick arrays of NPs deposited on substrates show larger values of $\zeta$ ranging from 2.2 to 3, which can be considered as intermediate between the 2D and the 3D case [27,28,43]. In our samples, as explained in III.B, only the $\zeta$ values extracted from samples B, D and E are valuable. They range from 3.5 to 5.2, *i.e.* much higher than what is reported in the 2D or 2D/3D cases. These high values may be due to the high dimensional nature of the current paths across the sample. The validation of this hypothesis deserves some numerical simulations aiming at calculating $\zeta$ in 3D arrays.

Except for sample X, there is no obvious relationship between the structural properties of the samples or the nature of the ligands and their electrical properties (room temperature resistance, $\zeta$ and $T_0$ values). We think that the fact that no clear Coulomb gap is observed in sample A and C simply results from the dispersion of the electrical properties between the different samples, and that a clear Coulomb gap could also be observed in these samples, but at lower temperatures. Indeed, as can be seen on sample B and D [see Figs. 5(a) and 5(c)], the opening of the Coulomb gap occurs sharply at low temperature in a very narrow temperature range.

The presence of hysteretic and abrupt transitions in the *I(V)* or in the *R(T)* characteristics of the samples, described in section III.C, has never been reported in arrays of chemically-prepared NPs. However, related behaviour has been reported in thin films of quenched condensed Al [31] and granular films of Cr [32]. One can also note the similarity between these phenomena and the electric-field induced switching of charge ordered-states observed in manganites [44,45,46,47]. In our system, the abrupt increase of the current and of the differential resistance means that several conduction channels open at the same time in the sample, which explains our appellation of "collective Coulomb blockade" for the observed phenomena. Numerical studies of transport in arrays of NPs have shown or mentioned that hysteresis [21], avalanches [30] or blocking in a metastable state [22] may occur when the capacitive coupling between neighbouring NPs $C$ is much larger than the self-capacitance $C_0$ of the NPs. Since $C_0$ scales as the diameter of the NPs and $C$ as their surface, arrays of large NPs should be more favourable to observe these phenomena. This may explain our results since our NPs were larger than the ones measured by other groups [7,24,43].

Several predictions by Sverdlov *et al.* on the electrical properties of strongly coupled NPs are consistent with our experimental results [30]. These predictions were i) an increase of electrical noise near the threshold voltage, which was indeed observed on sample A and C (see Fig. 6(b), 6(c) and Ref. [38]) ii) the possibility that infinite avalanches - *i.e.* an abrupt and irreversible increase of the current - occur in very large arrays iii) the fact that the avalanches are suppressed by thermal fluctuations. Experimentally, Fano factor measurements as a function of the temperature and of the applied voltage should be performed on these systems to study more deeply the nature of these transitions.

These results show that several theoretical questions are still opened concerning the transport in systems of strongly coupled magnetic NPs : i) the influence of the structural parameters (size, size dispersion, organisation, tunnel barrier thickness) and of the measurements conditions (applied voltage, temperature) on the properties of an eventual hysteresis ii) the relationship between the "Coulomb glass" state of electrons supposed to occur in systems of interacting localised electrons [29] and these particular transport properties; indeed, we suspect that the presence of hysteresis and avalanches might be a consequence of the formation of a Coulomb glass in the samples iii) the nature of the coupling between Coulomb blockade and magnetic-field.

We now discuss the high-field MR observed in the samples above 1.8 K and presented in section III.D. Given its huge amplitude, this high-field MR cannot be attributed to the tunnel MR generally observed in arrays of NPs. Instead, it can be noted that this high-field MR displays two features typical of the colossal MR observed in some magnetic oxides: i) the MR amplitude grows up with decreasing temperature and abruptly collapses below a critical temperature [see Fig. 8(a)] ii) the amplitude can easily reach very high values and does not saturate at high magnetic field (see Fig. 8(d) and Ref. [39]). Colossal MR is due to the fact that the transfer probability of electrons between two nearest Mn sites strongly depends on the angle between their magnetic moments. In our case, we interpret the high-field MR as resulting from the presence of paramagnetic states localised at the surface or between the NPs. This hypothesis is re-enforced by the $H/T$ dependence of the MR ratio [see Fig. 8(d)]. These paramagnetic centres could result from the coordination of the ligands at the surface of the NPs [39]. Alternately, since the investigated samples are as-grown systems, without any purification procedure, some intermediate species corresponding to molecular complexes formed during the first stage of the precursor decomposition may be still embedded in the super-lattices [48]. We developed a phenomenological model where the tunnelling occurs via a paramagnetic centre, which is indeed able to reproduce the large amplitude of the MR ratio and its strong voltage dependence [40]. In this framework, the abrupt drop of the high-field MR below 1.8 K could be explained by a magnetic transition of these impurities to an ordered state. Below 1.8 K, the observation of inverse tunnel MR is coherent with this interpretation of a high-field MR due to impurities. Indeed, the transfer between two magnetic electrodes through an impurity inside the tunnel barrier is known to induce an inverse TMR [49,50]. As a consequence, above (below) 1.8 K, the electronic transport would occur through a paramagnetic (frozen) impurity, leading to a high-field (inverse tunnel) MR.

The weak amplitude of the inverse tunnel MR may have two origins. First, the process of spin dependent tunnelling through an impurity always lead to a strong decrease of the TMR ratio compared to the one with direct tunnelling. Second, as seen in Fig. 2(b), the remnant magnetization in these super-lattices is almost null, which is the sign that the dipolar coupling between the NPs is strong compared to their anisotropy [51]. As a consequence, it is likely that, in the super-lattices, the reversal of the magnetization occurs via the reversal of large domains composed of ferromagnetically coupled NPs. In this case, the mean angle between adjacent NPs will be quite small during the reversal, leading to a reduced TMR compared to a reversal of NPs without interaction.

Finally, the experimental results obtained on sample X were of great interest since this sample was composed of non-organized broadly-dispersed NPs, but presented the most impressive electrical properties: i) The $\zeta = 1$ value at 1.9 K means that all the conduction channels opened at the same time above the threshold voltage [see Fig. 4(f)]. ii) At high voltage, it behaves as a purely metallic sample [see Fig. 5(f)] iii) Below 1.8 K, this sample displayed the largest jumps in the *I(V)* characteristics and the largest influence of the magnetic field on the threshold voltage [see Fig. 7(b)]. This clearly shows that a perfect organization of the NPs inside the super-lattices is not required to observe the collective effects described in this article. Much more precise chemical and structural characterisations are necessary to identify the origin of these atypical behaviours.

V. Summary and outlook

In this article, we have described some new behaviours appearing in large super-lattices of CoFe NPs. Collective Coulomb blockade phenomena have been observed in these large arrays of NPs, which manifest through abrupt transitions from a resistive to a conductive state. Two original mechanisms of MR leading to huge MR ratios have been observed, one of them being related to the collective Coulomb effect. Further investigations on other systems are needed to identify precisely the origin of these two mechanisms. The high-field MR observed above 1.8 K is thought to be due to paramagnetic centres arising from ligand surface coordination or from molecular species trapped in the organic matrix. Understanding their nature requires a more precise control of both the surface chemistry of the NPs and the organic matrix. For instance, the volunteer and controlled incorporation of well-defined molecular species between the NPs will be done, as well as a change in the nature of the ligands surrounding the NPs. The collective effects are thought to have a pure electrostatic origin, and thus may be observed in super-lattices of non-magnetic NPs such as gold. We think that the measurements of NP arrays with high mutual capacitance, *i.e.* large NPs separated by thin tunnel barriers, could favour the observation of these phenomena. Enlightening theoretical papers have been devoted to the transport properties of arrays of NPs in the strong coupling limit [22,30] but specific studies on the irreversibility in *I*(*V*) characteristics and on the eventual transition to a Coulomb glass are still missing in these systems. Finally, the origin of the coupling between the magnetic field and Coulomb blockade remains an open question.

**Acknowledgements :**

We acknowledge C. Crouzet and J. Moreau for their building of the sample holder. We thank P. Seneor and A. Bernand-Montel for fruitful discussions.

**References :**

[*] Correspondence should be addressed to M. R. and J. C. (marc.respaud@insa-toulouse.fr and julian.carrey@insa-toulouse.fr)

[+] current address: Materials Science and Engineering, Korea University, Anam-Dong Seongbuk-Gu, Seoul 136-713, KOREA.

[1] Q. A. Pankhurst, J. Connolly, S. K. Jones and J. Dobson, J. Phys. D : Appl. Phys. **36**, R167 (2003)

[2] X. M. Yang, C. Liu, J. Ahner, J. Yu, T. Klemmer, E. Johns, D. Weller, J. Vac. Sci. Technol. **22**, 31 (2004)

[3] X. Waintal and O. Parcollet, Phys. Rev. Lett. **94**, 247206 (2005)

[4] M. B. A. Jalil and S. G. Tan, Phys. Rev. B **72**, 214417 (2005)

[5] P. Seneor, A. Bertrand-Mantel and F. Petroff, J. Phys.:Condens. Matter **19**, 165222 (2007)

[6] H. Shimada, K. Ono and Y. Ootuka, J. Appl. Phys. **93**, 8259 (2003)

[7] K. Yakushiji, S. Mitani, F. Ernult, K. Kakanashi, H. Fujimori, Phys. Rep. **451**, 1 (2007)

[8] J. Wunderlich et al., J. Magn. Magn. Mat. **310**, 1883 (2007)

[9] M. Respaud et al. , Phys. Rev. B **57**, 2925 (1998)

[10] O. Margeat, D. Ciuciulescu, C. Amiens, B Chaudret, P. Lecante, M. Respaud., Progress in solid state chemistry **33**, 71 (2005)

[11] E. V. Shevchenko, D.V. Talapin, N. A. Kotov, S. O'Brien and C. B. Murray, Nature **439**, 55 (2006)

[12] H. Fan *et al.*, Science **304**, 567 (2004).

[13] C. Desvaux *et al.*, Nature Material*s* **4**, 750 (2005)

[14] C. T. Black, C. B. Murray, R. L. Sandstrom and S. Sun, Science **290**, 1131 (2000)

[15] S. Jang, W. Kong, and H. Zeng, Phys. Rev. B **76**, 212403 (2007)

[16] J. Inoue and S. Maekawa, Phys. Rev. B **53**, R11927 (1996)

[17] J. Joshua Yang, C. Ji, Y. A. Chang, X. Ke and M. S. Rzchowski, Appl. Phys. Lett. **89**, 202502 (2006)

[18] B. Hackenbroich, H. Zare-Kolsaraki, and H. Micklitz, Appl. Phys. Lett. **81**, 514 (2002)

[19] O. Chayka, L. Kraus, P. Lobotka, V. Sechovsky, T. Kocourek, and M. Jelinek, J. Mag. Mag. Mat. **300**, 293 (2006)

[20] H. Zare-Kolsaraki and H. Micklitz, Phys. Rev. B **67**, 094433 (2003)

[21] A. A. Middleton and N. S. Wingreen, Phys. Rev. Lett. **71**, 3198 (1993)

[22] D. M. Kaplan, V. A. Sverdlov, and K. K. Likharev, Phys. Rev. B **68**, 045321 (2003)

[23] C. Reichhardt and C. J. O. Reichhardt, Phys. Rev. Lett. **90**, 046802 (2003)

[24] R. Parthasarathy, X.-M. Lin, K. Elteto, T. F. Rosenbaum and H. M. Jaeger, Phys. Rev. Lett. **92**, 076801 (2004)

[25] K.-H. Müller *et al.* , Phys. Rev. B **66**, 075417 (2002)

[26] C. Kurdak, A. J. Rimberg, T. R. Ho, J. Clarke, Phys. Rev. B **57**, R6842 (1998)

[27] H. Fan et al. Science **304**, 567 (2004)

[28] P. Beecher, A. J. Quinn, E. V. Shevchenko, H. Weller and G. Redmond, Nanoletters **4**, 1289 (2004)

[29] S. Pankov and V. Dobrosavljević, Phys. Rev. Lett. **94**, 046402 (2005)

[30] V. A. Sverdlov, D. M. Kaplan, A. N. Korotkov, K. K. Likharev, Phys. Rev. B **64**, 041302(R) (2001)

[31] E. Bielejec and W. Wu, Phys. Rev. Lett. **87**, 256601 (2001)

[32] V. A. Krupenin, V. O. Zalunin and A. B. Zorin, Microeletron. Eng. **81**, 217 (2005)

[33] E. Lebannon and M. Müller, Phys. Rev. B **72**, 174202 (2005)

[34] K. Yakiushiji, S. Mitani, F. Ernult, K. Takanashi and H. Fujimori, Physics Report **451**, 1 (2007)

[35] S. Mitani *et al.*, Phys. Rev. Lett. **81**, 2799 (1998)

[36] S. J. van der Molen, N. Tombros and B. J. van Wees, Phys. Rev. B **73**, 220406 (R) (2006)

[37] A. Bernand-Mantel, P. Seneor, R. Guillemet, S. Fusil, K. Bouzehouane, C. Deranlot, F. Petroff and A. Fert, submitted

[38] R. P. Tan et al., Phys. Rev. Lett. **99**, 176805 (2007)

[39] R. P. Tan et al., J. Magn. Magn. Mater. **320**, L55 (2008)

[40] R. P. Tan, J. Carrey and M. Respaud, J. Appl. Phys. **104**, 023908 (2008)

[41] R. P. Tan et al., J. Appl. Phys **103**, 07F317 (2008)

[42] C. Desvaux, F. Dumestre, C. Amiens, M. Respaud, P. Lecante, E. Snoeck, P. Fejes, P. Renaud and B. Chaudret, submitted

[43] C. Lebreton, C. Vieu, A. Pépin, M. Mejias, F. Carcenac, Y. Jin and H. Launois, Microelec. Engin. 41/42, 507 (1988)

[44] A. Asamitsu, Y. Tomoioka, H. Kuwahara and Y. Tokura, Nature **388**, 50 (1997)

[45] K. Hatsuda, T. Kimura and Y. Tokura, Appl. Phys. Lett. **83**, 3329 (2003)

[46] Y. Tokura and Y. Tomioka, J. Magn. Magn. Mater. **200**, 1(1999)

[47] H. Kuwahara, Y. Tomioka, A. Asmitsu, Y. Moritomo and Y. Tokura, Science **270**, 961 (1995)

[48] L.-M. Lacroix, S. Lachaize, A. Falqui, M. Respaud, B. Chaudret, J. Am. Chem. Soc., in press

[49] E. Y. Tsymbal, A. Sokolov, I. F. Sabirianov and B. Doudin, Phys. Rev. Lett. **90**, 186602 (2003)

[50] J. R. Petta, S. K. Slater, and D. C. Ralph, Phys. Rev. Lett. **93**, 136601 (2004)

[51] D. Kechrakos and K. N. Trohidou, Phys. Rev. B **58**, 12169 (1998)

**Table**

| Sample (NPs size) | Ligands | R (300 K) | $T_0$ | $\zeta$ | $V_T$ | high-field MR | avalanches |
|---|---|---|---|---|---|---|---|
| A (25 nm) | 2 OA | 400 kΩ | 24 K (2 V) | 2.5 (b) | 0 (b) | not measured | Yes (36%) |
| B [39] (15 nm) | 1 OA + 1 SA | 5 GΩ | 80 K (25 V) | 3.5 (fixed) 4.37 | 33 0(fixed) | 3000 % | No |
| C [38] (15 nm) | 1 OA + 1 SA | 8 kΩ | 24 K (10 mV) | 1.3 (b) | 2 (b) | 100 % | Yes (120%) |
| D (15 nm) | 1 OA | 100 MΩ | 256 K (10 V) | 5.2(fixed) 4.4 | 3.2 30(fixed) | 90 % | No |
| E (25 nm) | 2 OA | 1.2 GΩ | 244 K (1 V) | 4.9 (fixed) 4.03 | 24 50(fixed) | 160 % | No |
| X (25 nm) | 1 OA + 1 SA (c) | 300 kΩ | 200 K (20 mV) | 1 (a) | 30 | 280 % | Yes (3000%) |

Table I :

Summary of the super-lattices properties. For sample B and C, references of articles in which some of their properties have been previously published are indicated in the first column. The column "ligands" corresponds to the amount and the nature of acids and amines used for the synthesis. The subsequent columns display the values of the resistance at room temperature $R$ (300 K), the activation energy $T_0$, the exponent $\zeta$ and the threshold voltage $V_T$. In the column "$T_0$", the voltage at which the $R(T)$ has been measured is indicated in parentheses. The column "high-field MR" indicates the maximum high-field MR measured on the sample. The last column indicates if avalanches have been observed in the $I(V)$ characteristics at low temperature, and provides in parentheses the amplitude of the avalanches. (a) On sample X, the $I(V)$ characteristics are linear as soon as the voltage is above $V_T$ (see text and Fig. 4(f)). (b) On these samples, the gap in the $I(V)$ characteristic at 1.8 K is not clearly defined. (c) For sample X, there was a problem during the synthesis which manifest through a large size distribution (see text, and Fig. 1).

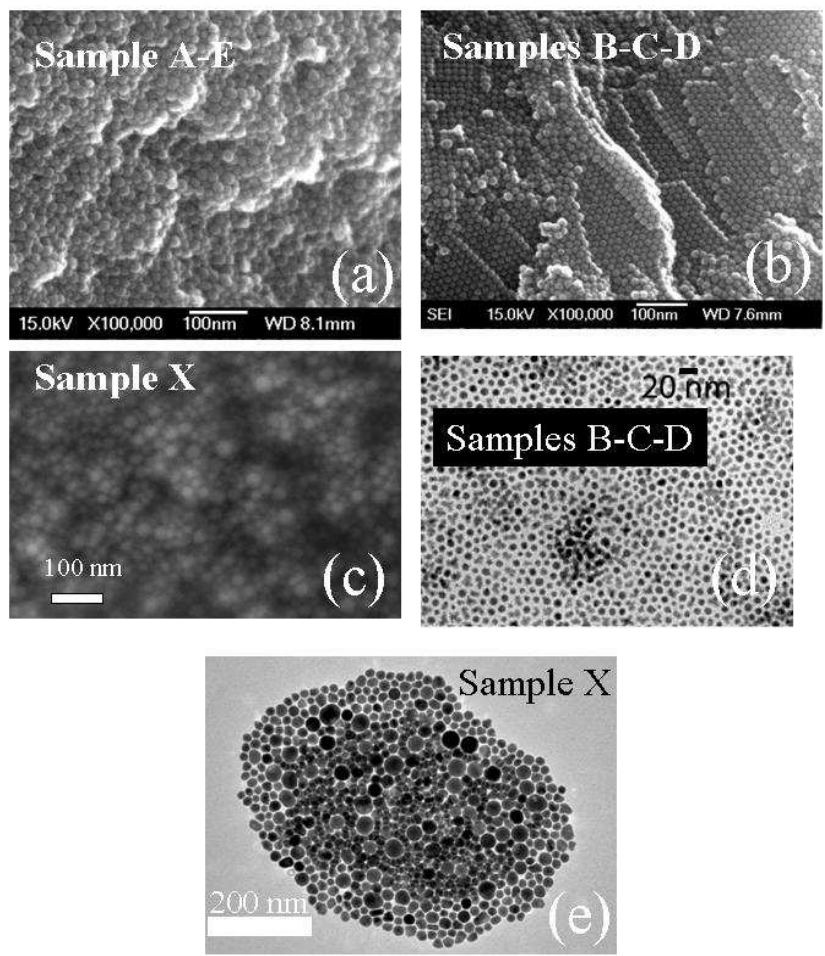

Fig. 1

Fig. 1: a) b) c) SEM micrographs of super-lattices synthesized in various conditions (see table I). The micrographs illustrate the particles size and organisation for a) sample A; b) sample B, C and D; c) Sample X. d) e) TEM micrographs of the NPs after dispersion of the super-lattices in solvent for sample B, C and D (d), and sample X (e).

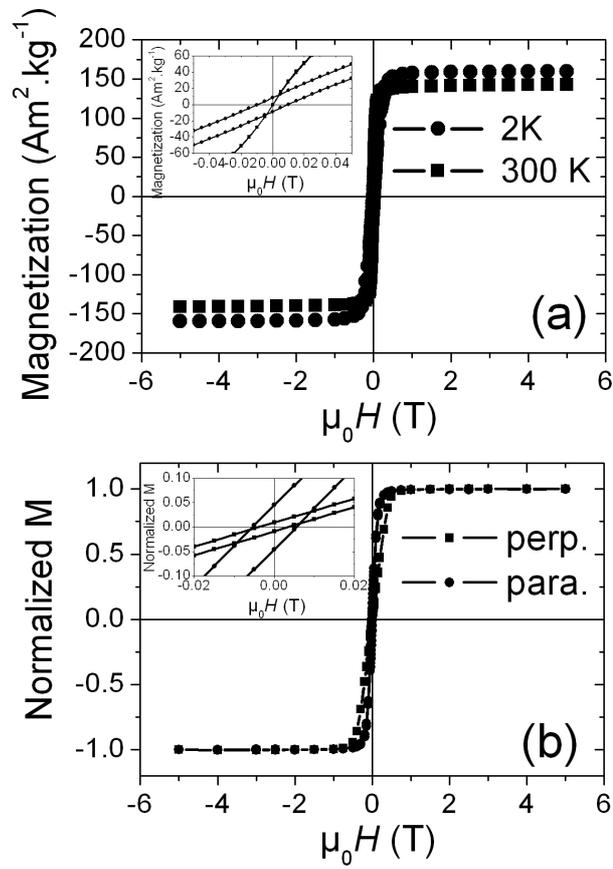

Fig. 2

Fig. 2: a) *M*(*H*) measured at 2 K and 300 K on a powder composed of a large number super-lattices. b) *M*(*H*) of a single super-lattice measured with the field perpendicular or parallel to the longest axis of the super-lattice. (insets) Enlarged view of the *M*(*H*)s at low magnetic field.

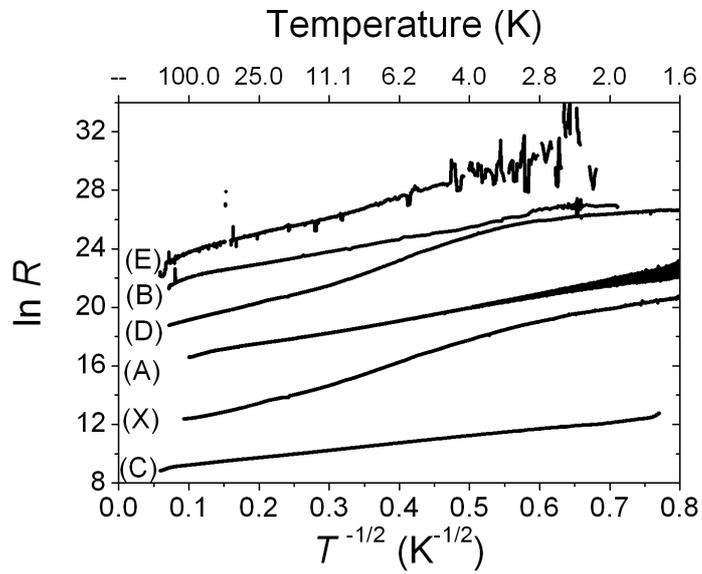

Fig. 3: Resistance of the different samples as a function of temperature. The logarithm of the resistance is plotted versus $T^{-1/2}$.

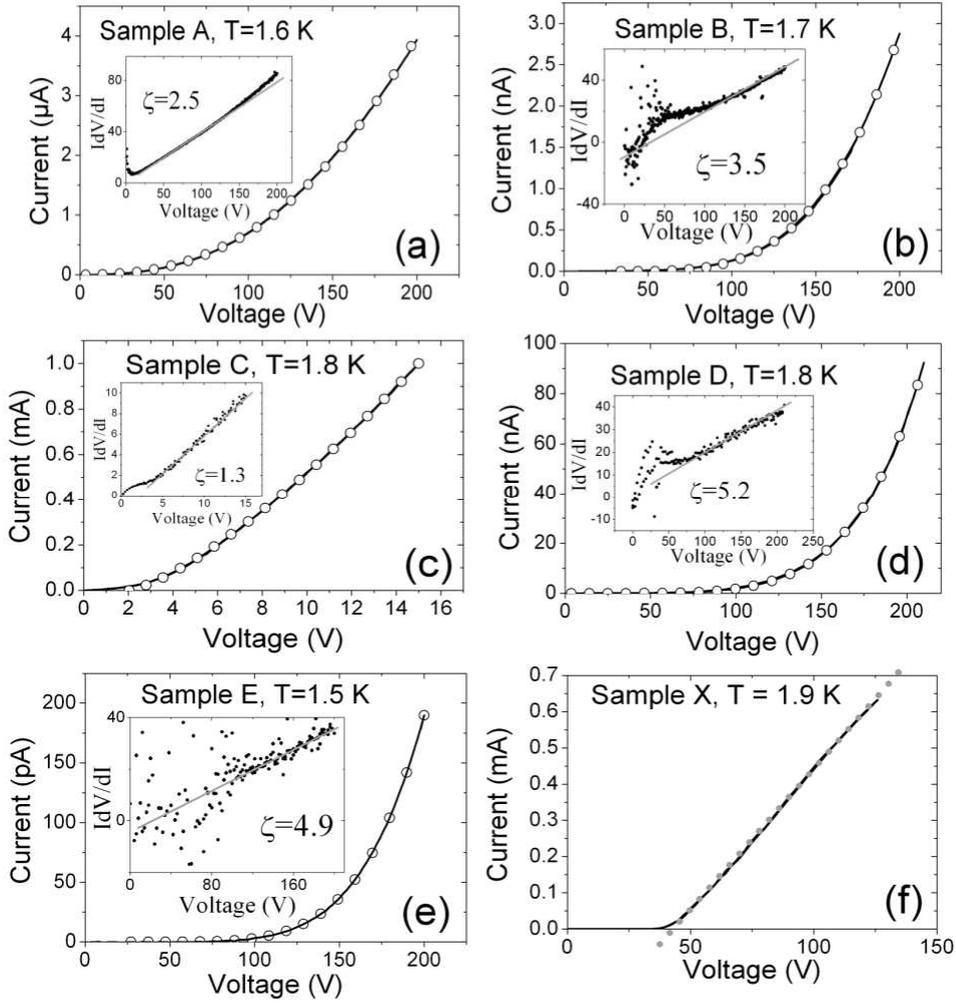

# Fig. 4

Fig. 4: a) b) c) d) e) $I(V)$ characteristics of the super-lattices (straight lines) and the corresponding fits using Equ. (2) (open circles). The measurement temperatures are indicated on the figures. The $\zeta$ values used for the fits are deduced from the plot of $IdV/dI$ versus $V$, shown in the inset. The $V_T$ values deduced from the best fits of the experimental data are given in Table I. f) $I(V)$ characteristic at 1.9 K of sample X (straight line). A linear slope is shown as a guide to the eye (dotted line).

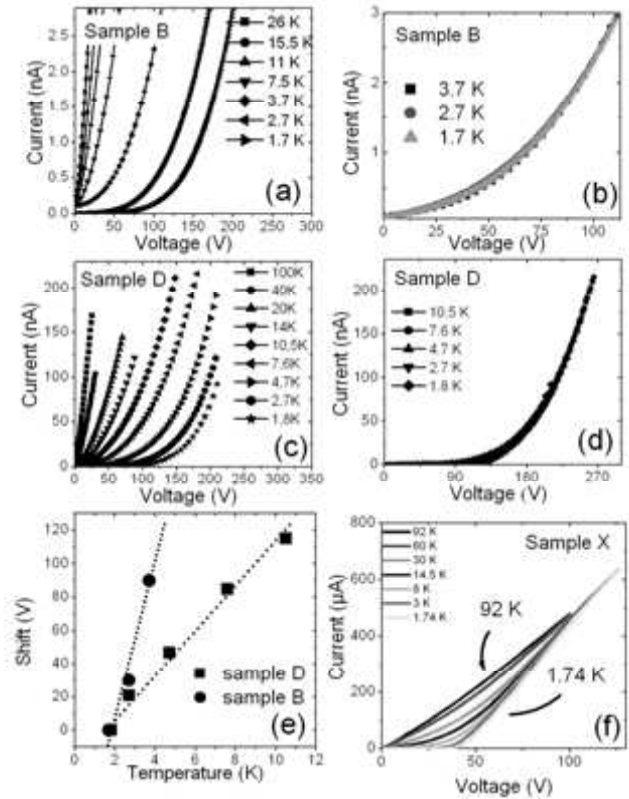

# Fig. 5

Fig. 5 (color online): a) c) Evolution of the *I*(*V*)s for samples B and D as a function of temperature. b) d) *I*(*V*)s at different temperatures for samples B and D are shifted to superpose. e) Plot of the shift values as a function of temperature for samples B and D. Dotted lines are a guide to the eye. f) Evolution of the *I*(*V*)s as a function of the temperature for sample X.

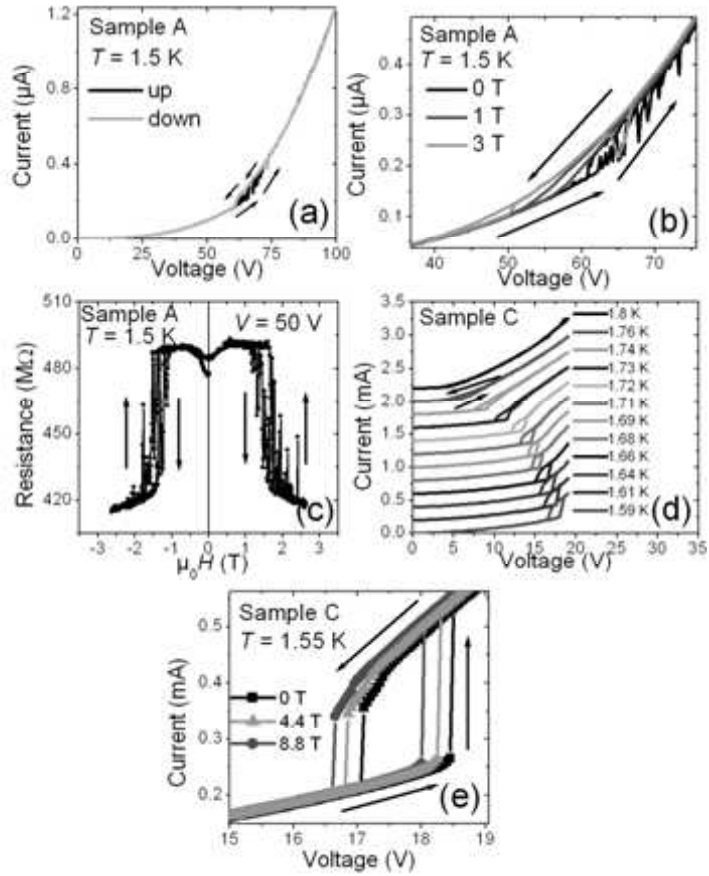

**Fig. 6:**

Fig. 6 (color online): a) I(V) of sample A at 1.5 K. The arrows indicate the voltage sweep direction. b) Evolution of the I(V) of sample A at 1.5 K under the application of a magnetic field of 0, 1 and 3 T. c) R(H) of sample A at 1.5 K with a voltage bias of 50 V. The arrows indicate the magnetic field sweep direction. d) Evolution of the I(V) of sample C as a function of temperature. e) Evolution of the I(V) of sample C at 1.55 K under the application of a magnetic field of 0, 4.4 and 8.8 T.

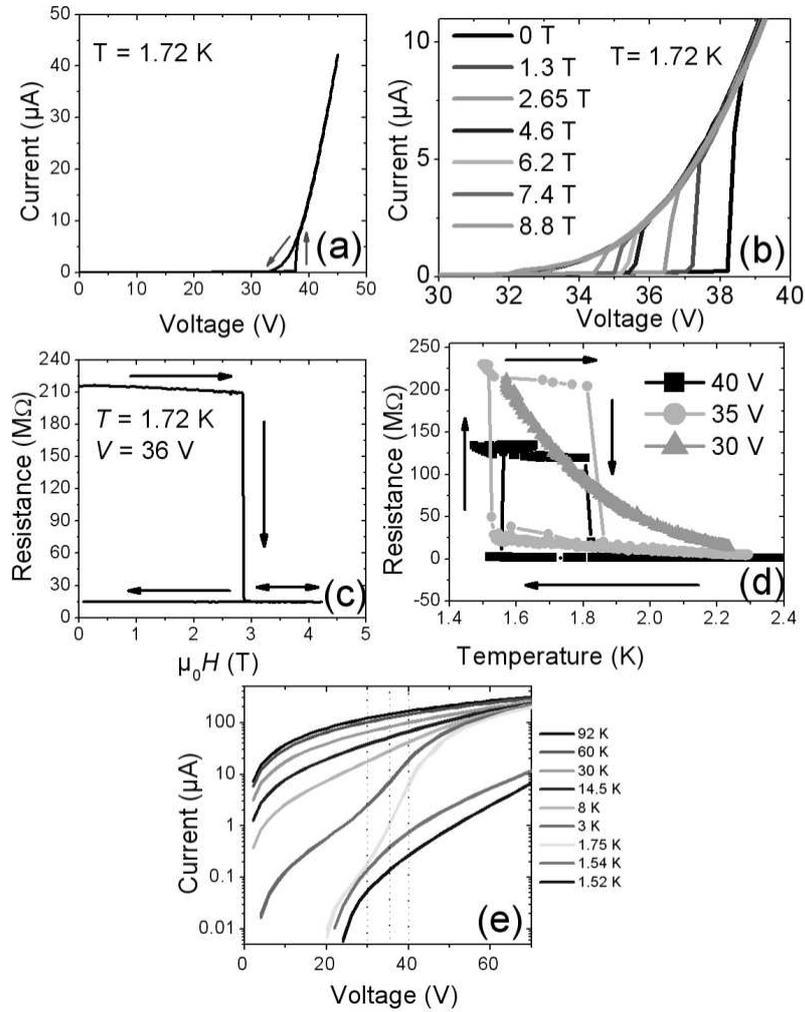

**Fig. 7:**

Fig. 7 (color online): Transport properties of sample X. a) $I(V)$ at $T$=1.72 K. b) $I(V)$ characteristics at 1.72 K and different values of the magnetic field. The sweep direction is similar to (a). c) $R(H)$ measurement at $T = 1.72$ K and $V = 36$ V. Arrows indicate the sweep direction. d) $R(T)$ measured at 30 V, 35 V and 40 V. Arrows indicate the sweep direction. For the measurement at 30 V, the $R(T)$ does not display any hysteresis. e) $I(V)$s at different temperatures on a semi-logarithmic scale.

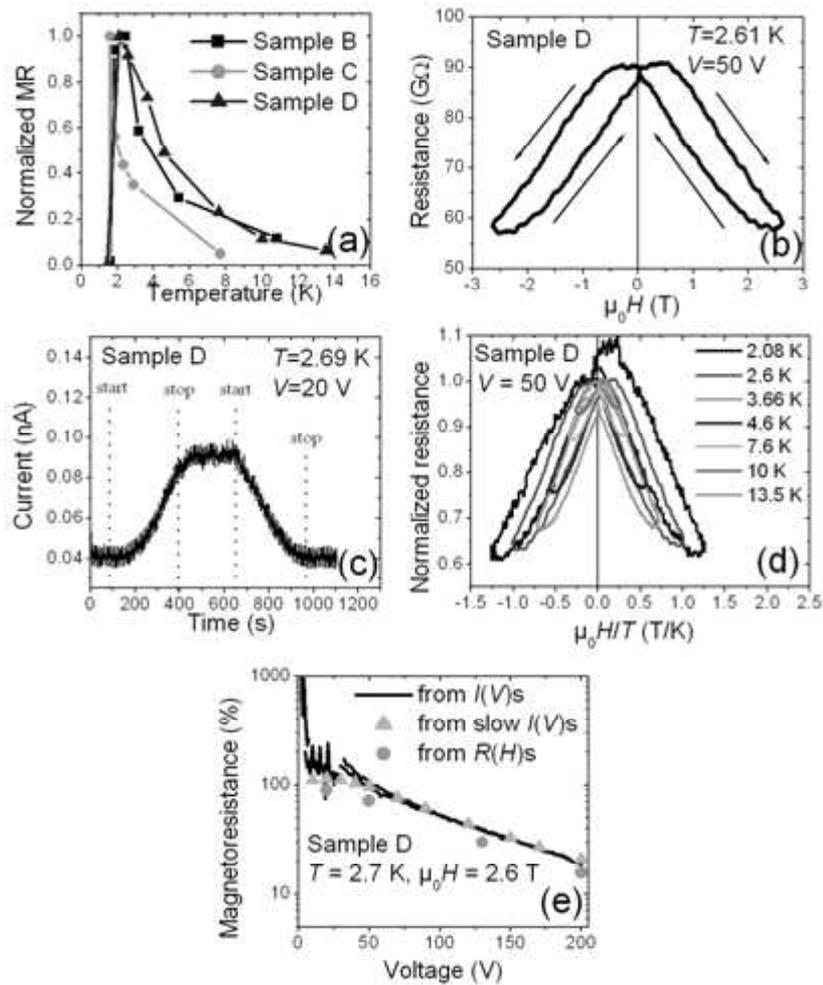

**Fig. 8**

Fig. 8 (color online): a) MR(*T*) for samples B, C and D. The MR amplitudes were deduced from *R*(*H*) measurements at 2.6 T and were normalized to the highest value (17 %, 80 % and 64 %, respectively). The applied voltage was 200 V, 20 mV and 50 V, respectively. b) *R*(*H*) on sample D measured at *T* = 2.61 K and *V* = 50 V. c) Evolution as a function of time of the current in sample D when the magnetic field is swept from 0 to 2.6 T and then back. The "start" ("stop") corresponds to the start (end) of the magnetic field variation. d) *R*(*H*/*T*) characteristics for sample

D measured at various temperatures with $V = 50$ V. The resistance has been normalized to its value at zero magnetic field. e) MR($V$) at $\mu_0 H = 2.6$ T for sample D at 2.7 K. Orange dots corresponds to values deduced from $R(H)$ measurements at different voltages. Black line and blue triangles correspond to the values deduced from two $I(V)$ characteristics at zero field and 2.6 T. $I(V)$s corresponding to the blue triangles have been manually measured using the best measurement range for each point.

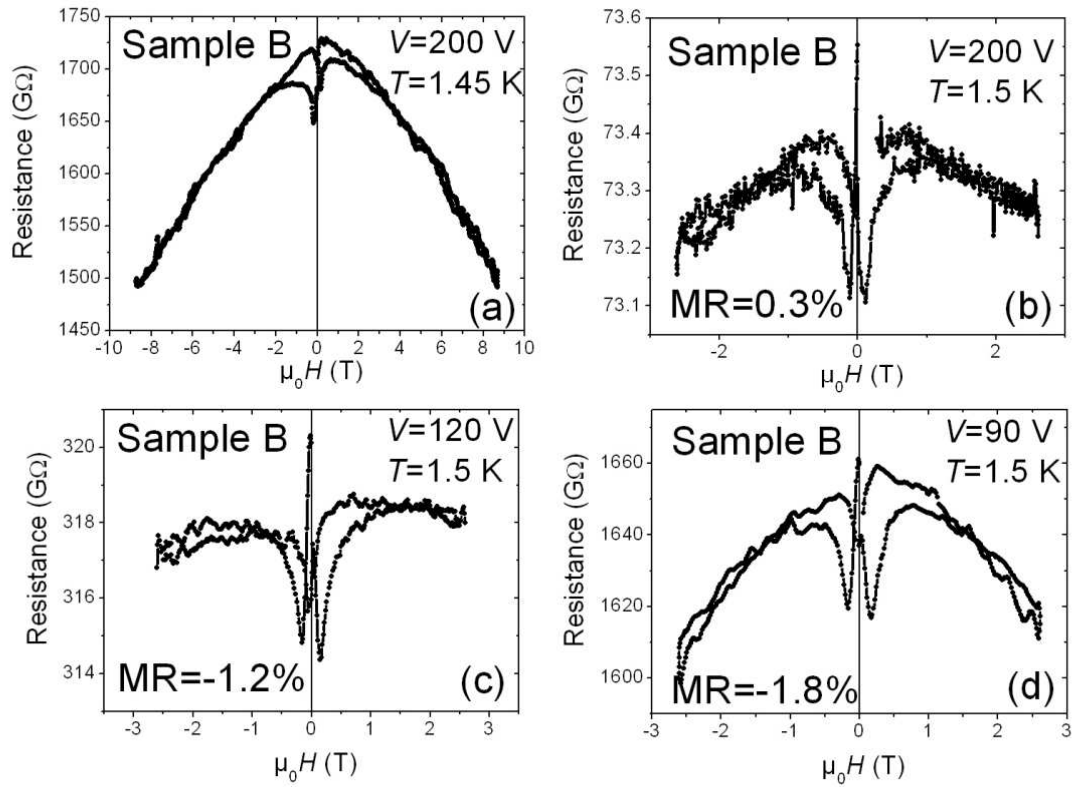

Fig. 9

Fig. 9: MR of sample B below 1.8 K. The MR value indicated on the figures is taken between the inverse peaks and the plateau at 1 T. Fig. a) is measured at a maximum magnetic field of 8.5 T and 1.45 K, the other ones at 2.5 T and 1.5 K. a) and b) voltage bias $V$=200 V. c) $V$=120 V. d) $V$ = 90 V.

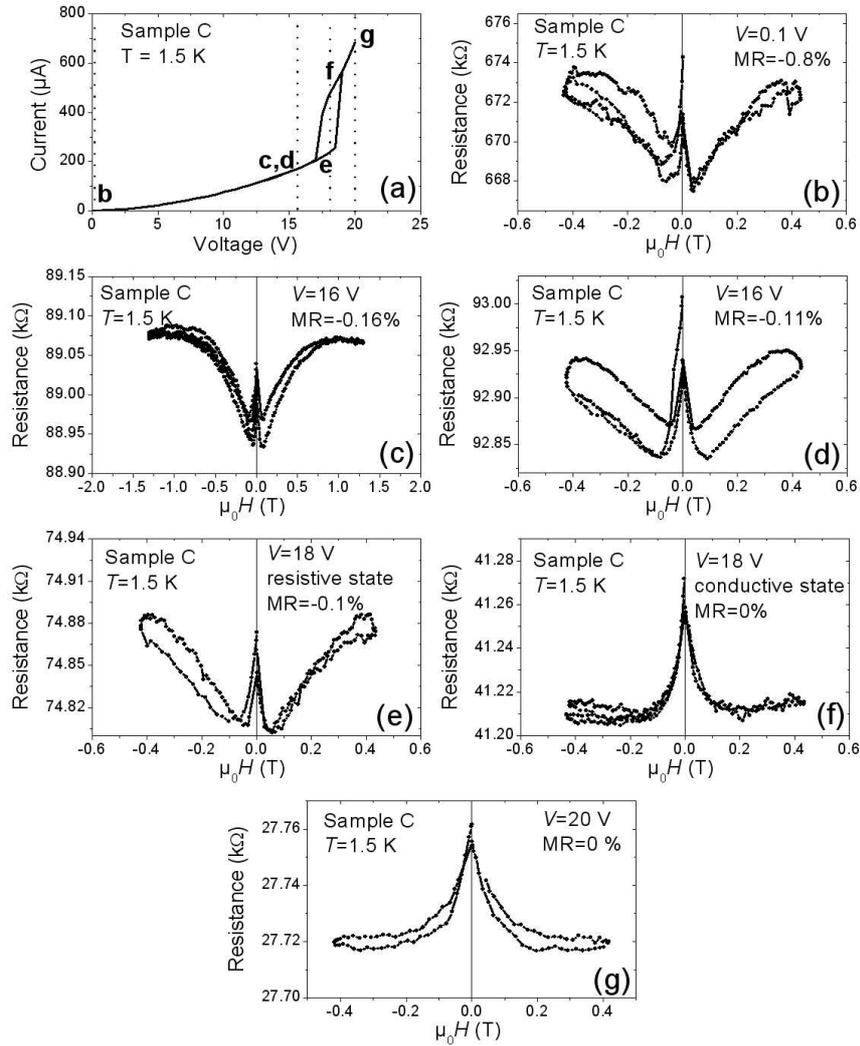

Fig. 10 : a) *I(V)* of sample C at 1.5 K. The dotted line indicates the voltages at which the MR measurements have been performed. b) c) d) e) f) g) MR measurements performed on sample C at 1.5 K. The bias voltage at which the MR have been performed are indicated on the figures. For a bias voltage of 18 V, there are two possible states of conduction. For e) the sample is in the resistive state; for f) it is in the conductive state. The MR amplitudes indicated on the graphs do not take into account the central peak.